\documentclass[aps,prb,twocolumn,floats]{revtex4}
\usepackage{epsfig}
\usepackage{amsmath}
\usepackage{graphicx}
\usepackage{color}

\begin{document}
\title{Fermi Level Fluctuations, Reduced Effective Masses and Zeeman Effect during Quantum Oscillations in Nodal Line Semimetals}
\author{Satyaki Kar$^*$, Anupam Saha}
\affiliation{AKPC Mahavidyalaya, Bengai, West Bengal -712611, India}
\begin{abstract}
  We probe quantum oscillations in nodal line semimetals (NLSM) by considering a NLSM continuum model under strong magnetic field and report the characteristics of the Landau level spectra and the fluctuations in the Fermi level as the field in a direction perpendicular to the nodal plane is varied through. Based on the results on parallel magnetization, we demonstrate the growth of quantum oscillation with field strength as well as its constancy in period when plotted against 1/B. We find that the density of states which show series of peaks in succession, witness bifurcation of those peaks due to Zeeman effect. For field normal to nodal plane, such bifurcations are discernible only if the electron effective mass is considerably smaller than its free value, which usually happens in these systems. Though a reduced effective mass $m^*$ causes the Zeeman splitting to become small compared to Landau level spacings, experimental results indicate a manyfold increase in the Lande $g$ factor which again amplifies the Zeeman contribution.
  We also consider magnetic field in the nodal plane for which the density of state peaks do not repeat periodically with energy anymore. The spectra become more spread out and the Zeeman splittings become less prominent. We find the low energy topological regime, that appears with such in-plane field set up, to shrink further with reduced $m^*$ values. However, such topological regime can be stretched out in case there are smaller Fermi velocities for electrons in the direction normal to the nodal plane.
\end{abstract}

\maketitle                              
\section{Introduction}
Nodal line semimetals\cite{burkov,fang,rev1} (NLSM) have become popular candidate among topological semimetals\cite{bernevig} due to their exotic features that are rapidly getting exposed in recent times. Its easy tunability to convert into other exotic materials such as a Weyl semimetal\cite{shen,ashvin,kar1}, Dirac semimetals\cite{shen,ashvin}, magnetic semiconductors\cite{ashvin} etc has also made these compounds interesting to the condensed matter community. A NLSM is characterized by topologically robust nodal ring or loops where the conduction and valence bands meet.
Such nodal loops, however, do not appear necessarily at equal energies as there are both type-I and type-II NLSM materials possible\cite{he,lim2}. Landau quantization develops under the influence of strong magnetic fields in such systems\cite{lim2,cortijo,kar2} which substantially changes the spectral features.
With field perpendicular to the nodal plane one can analytically obtain the Landau quantized spectra starting from naive continuum models of such systems. However a field in the nodal plane disrupts the planar symmetry and one needs to resort to numerics to deal with such problems.
Here we consider both the situations for a type I NLSM ssytem and try to estimate various factors related to quantum oscillations that results when the magnetic field strength is varied through.

The Landau tube widths and hence the density of states (DOS) profile changes continually with magnetic field. It causes the chemical potential ($\mu$) to experience oscillations as the field is varied through. Not only just $\mu$, oscillations appear in many physical variables such as magnetization, magnetoresistance or conductivity and such responses have been reported from various NLSM compounds\cite{zrsis,caagas,hfsis}. Theoretically, if Zeeman field term is considered in addition, splitting of the DOS peaks are obtained, though it can be noticed if the electronic effective mass is considerably smaller than the free electron mass as the Landau level (LL) spacing become much larger compared to the Zeeman splitting under such scenario.

Now such magnetic oscillations can be topological or trivial depending on the direction of the field applied and in fact nontrivial topology is obtained when the magnetic field contain nonzero components in the nodal plane\cite{cortijo,li,lim2}.
In this regard, here we also address the spectral features and their connection to topology as the field direction is confined in the nodal plane.
The recent theoretical studies on quantum oscillations in NLSM systems does not discuss much about Zeeman splitting in the spectra or in general the variation of effective masses ($m^*$) as a magnetic field is applied, even though experimentally the Zeeman splittings are well observed at low temperatures, high fields as well as low $m^*$ values. Here in this article we would like to theoretically examine that aspect of the problem and look for any interesting physics that it leads to.

The paper is organized as follows. In section II, we give the formulation for NLSM model under the influence of magnetic field perpendicular to the nodal plane.
In section III, we discuss the Fermi level fluctuation and quantum oscillations in magnetization. Section IV deals with effective mass reduction, Zeeman contributions whereas section V describes the density of states. In section VI we consider the field direction to be in the nodal plane and briefly repeat the earlier steps for this scenario. Finally we summarize our work in section VII.


\section{Spectral Formulation}
{As a NLSM features band crossings in a closed loop of states in the three-dimensional Brilluin zone (BZ)\cite{weng}, a typical} simple NLSM Hamiltonian\cite{photo,molina} (type-I) looks like
\begin{eqnarray}
H_0=(\frac{\hbar^2k_\perp^2}{2m^\star}-{\Delta})\sigma_z+v\hbar k_z\sigma_y
\label{eq1}
\end{eqnarray}
where $k_\perp^2=k_x^2+k_y^2$ and $m^\star$ denotes the effective mass of the band electrons. {The $\sigma~(\tau)$ matrices are the Pauli matrices describing orbital (spin) degrees of freedom and $v$ denotes the Fermi velocity of band electrons. Here $k_\perp^2=\frac{2{\Delta} m^\star}{\hbar^2}$ in the $k_z=0$ plane denotes the nodal loop/ring.}

In presence of a magnetic field perpendicular to the nodal plane in a nodal line semimetal, the Zeeman term or the coupling of spins with the field result in shift of nodal loops to nonzero energies. The Hamiltonian becomes

\begin{eqnarray}
H=(\frac{\hbar^2{k'}_\perp^2}{2m^\star}-{\Delta})\sigma_z\tau_0+v\hbar k'_z\sigma_y\tau_0+b\sigma_0\tau_z
\label{eq2}
\end{eqnarray}
where the last term represent the Zeeman interaction with $b=g\mu_BB/2$ and Peierl's substituted momentum ${\bf k'=k-eA/\hbar}$. The energy spectrum in presence of the magnetic field is given as
\begin{equation}
  \epsilon=\pm\sqrt{[\frac{\hbar^2{k'}_\perp^2}{2m^\star}-{\Delta}]^2+v^2\hbar^2{k'}_z^2}\pm b.
\end{equation}
Note that even in presence of the Zeeman term, the extremal orbits occur at $k'_z=0$.

On applying a magnetic field ${\bf B}=(0,0,B)$ perpendicular to the nodal loop, the modified Hamiltonian can be obtained with vector potential ${\bf A}$ expressed in the Landau gauge as ${\bf A}=(-By,0,0)$. One can rewrite the Hamiltonian as
\begin{eqnarray}
H(B)&=&H_0+\frac{e\hbar}{m^*}(-k_xBy+\frac{eB^2y^2}{2\hbar})\sigma_z\tau_0+b\sigma_0\tau_z
\end{eqnarray}
which, in the basis of Landau states\cite{kar2,molina,fazekas}, is written as
\begin{eqnarray}
H(B)=[(n+\frac{1}{2})\frac{e\hbar B}{m^*}-{\Delta}]\sigma_z\tau_0+v\hbar k_z\sigma_y\tau_0+b\sigma_0\tau_z.
\label{eqLL}
\end{eqnarray} 
So the dispersions take the form
\begin{eqnarray}
  \epsilon_{n,k_z}=\pm\sqrt{[(n+\frac{1}{2})\frac{e\hbar B}{m^*}-{\Delta}]^2+v^2\hbar^2k_z^2}\pm b .
  \label{eq-en}
\end{eqnarray}
Since $k_z$ is not affected by ${\bf B}$, the 3D problem decomposes into a family of 2D ones parametrized by $k_z$\cite{lim1}.
And for each $k_z$, we find discrete Landau level states (given by index $n$) with huge degeneracy that is also proportional to the magnetic field strength B\cite{fazekas}.

\section{Quantum Oscillations}
As a magnetic field variation causes Landau tubes to change their widths as well as allow them to successively cross through the Fermi surface (FS), the Fermi energy (or chemical potential $\mu$) or parallel magnetization $M_z$ quite naturally show fluctuations with such field variations. For now, here we start with $m^*=m$, the free electronic mass.

\begin{figure}
\includegraphics[width=\linewidth,height=2.5 in]{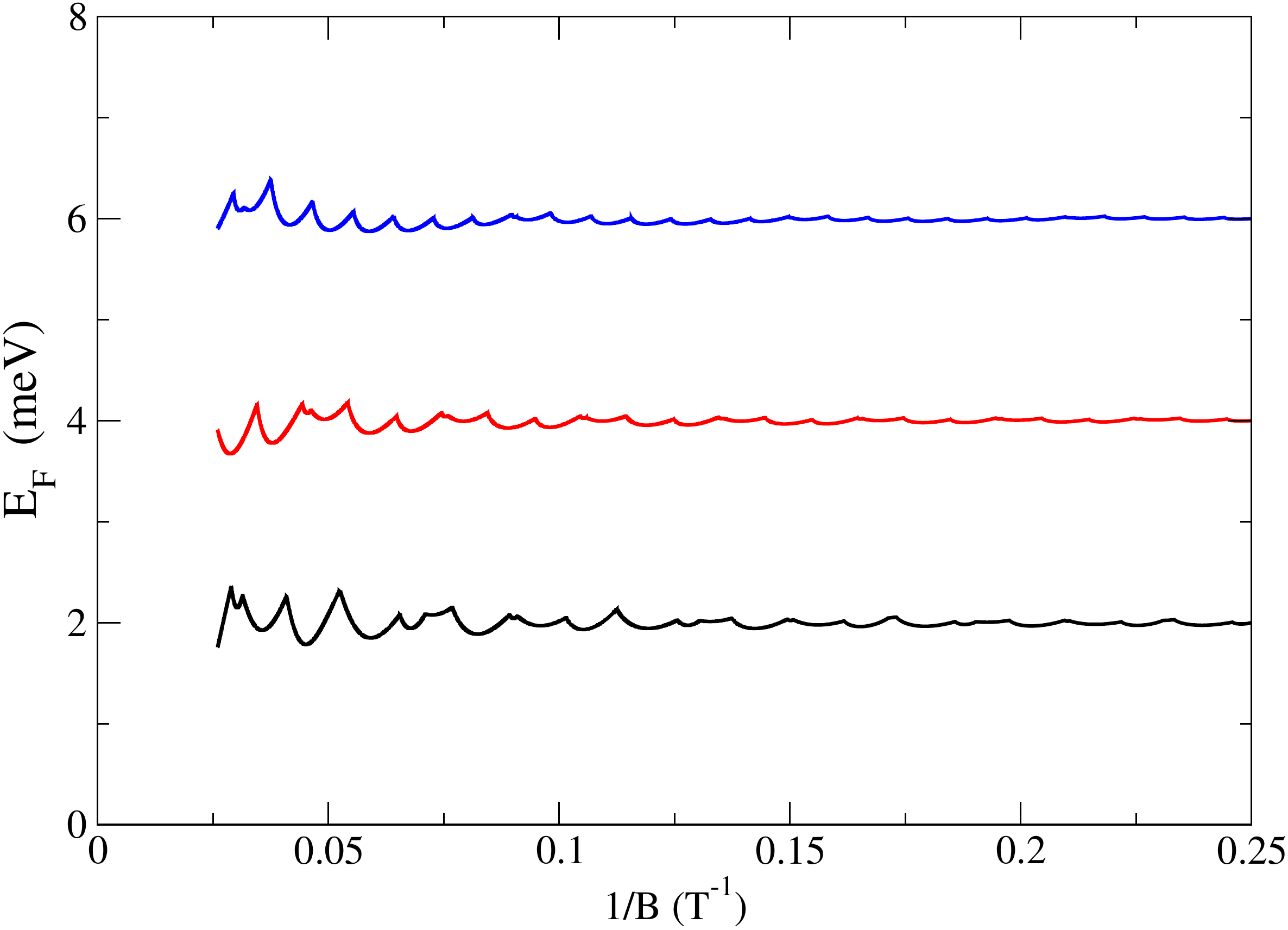}
\caption{Fluctuation in Fermi level plotted against inverse of field strength $B^{-1}$ for three different $N$ values.} 
\label{vary-mu}
\end{figure}

One can calculate the magnetization or total number of particles/electrons $N$ from the free energy $F=E-TS$ and thermodynamic potential $\Omega=F-\mu N$. In this paper we present a zero temperature analysis\cite{shoen} where one can write the contribution $\delta\Omega$ of the thermodynamic potential due to a small differential $\delta k_z$ to be
\begin{eqnarray}
  \delta\Omega=D\sum_{\epsilon_n\le\mu}(\epsilon_n-\mu)
  \label{dOmega}
\end{eqnarray}
with the sum over Landau level index is restricted as shown. Here $D$ is the degeneracy factor and is given by $D=\delta k_z \frac{eBV}{2\pi^2\hbar},~V$ being the volume of the sample. From $\Omega$ one can obtain $M_z$ and $N$ as $M_z=-\left(\frac{\partial\Omega}{\partial B}\right)_\mu$ and $N=-\left(\frac{\partial\Omega}{\partial\mu}\right)_B$.
If we disregard the Zeeman term (just for ease of calculation to start with) and  integrate  Eq.\ref{dOmega}, one can obtain the thermodynamic potential as
\begin{eqnarray}
 \Omega=&\frac{eBV}{2\pi^2\hbar^2v}\sum_n\{\frac{x_n^2}{2}log|\frac{\mu+y_n}{\mu-y_n}|-\mu y_n\}.
\end{eqnarray}
The sum includes those integers $n$ for which $y_n^2=\mu^2-[(n+1/2)\mu_BB-{\Delta}]^2=\mu^2-x_n^2\ge0$.
From there one gets the expression for $N$ as
\begin{eqnarray}
  N=\frac{eBV}{\pi^2\hbar^2v}\sum_{|\mu|>|x_n|}\sqrt{\mu^2-[(n+1/2)\mu_BB-{\Delta}]^2}.
\end{eqnarray}
Considering constancy of $N$ with field variation, one can numerically obtain the variation of Fermi energy $E_F$ (or $\mu$) with $B$ as shown in Fig.\ref{vary-mu}.
There are quite a few things that this plot indicates. First of all, quantum oscillation dominates at large $B$ values. The oscillation shows a constant periodicity when plotted against inverse of field. Thirdly, we notice huge fluctuations at very large $B$ values because very few Landau tubes can remain within the Fermi surface chosen here.

\begin{figure}
\includegraphics[width=\linewidth,height=2.5 in]{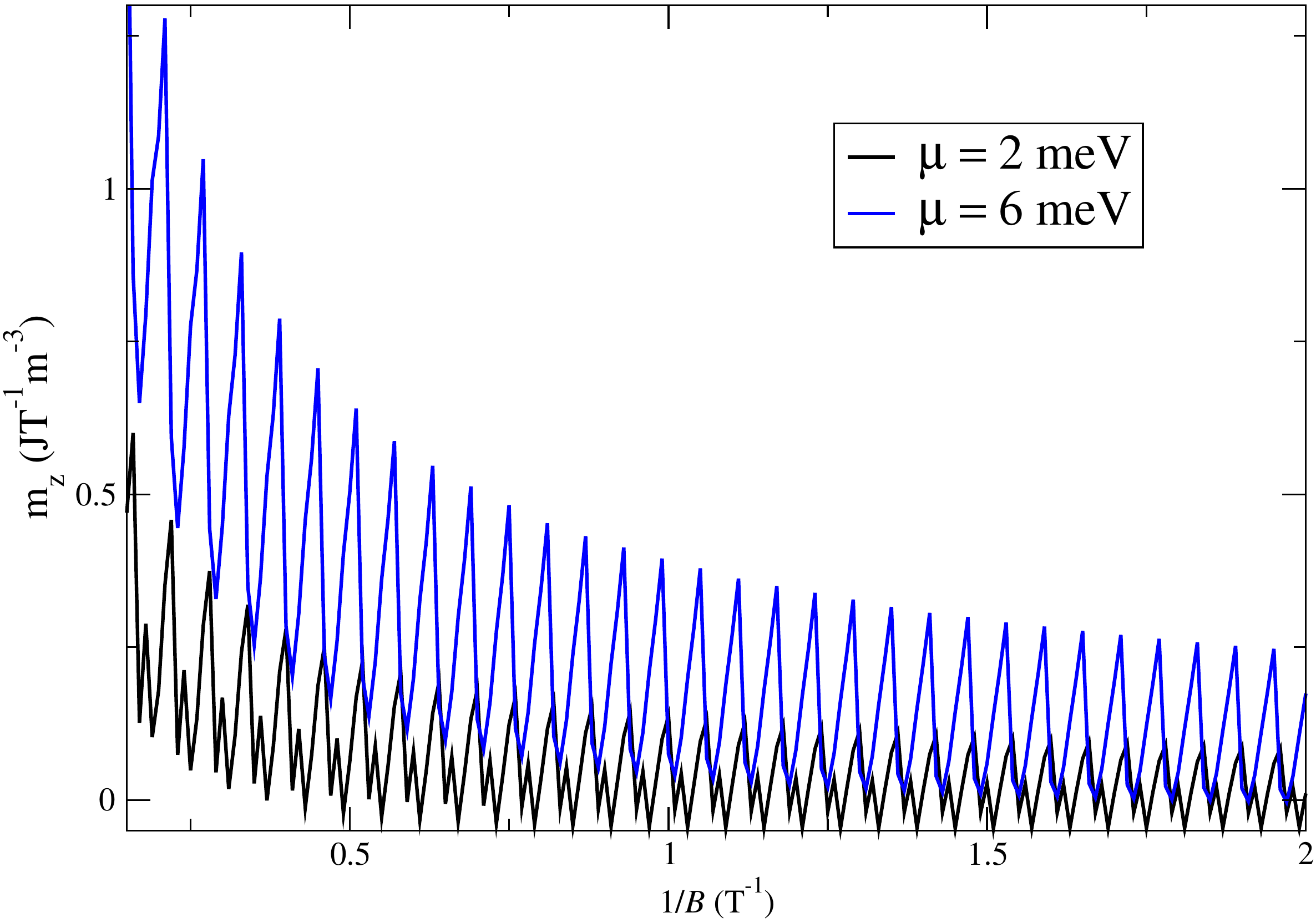}
\caption{Variation of $m_z$ with inverse of field strength $B^{-1}$ for different(average) $\mu$ values and ${\Delta}$=8 meV.} 
\label{vary-mz}
\end{figure}
Next we consider $M_z$, the magnetization parallel to field direction.
We switch on the Zeeman term and with that the $M_z$ expression  becomes
\begin{align}
  M_z=&\frac{eV}{2\pi^2\hbar^2v}[\sum_{|\mu_+|>|x_n|}\{\mu_+ y_n-x_n(\frac{3x_n}{2}+{\Delta})ln\frac{\mu_++y_n}{\mu_+-y_n}\nonumber\\
    &+2by_n\}+\sum_{|\mu_-|>|x_m|}\{\mu_- y_m-\nonumber\\
    &x_m(\frac{3x_m}{2}+{\Delta})ln\frac{\mu_-+y_m}{\mu_--y_m}-2by_m\}]
\end{align}
where $\mu_\pm=\mu\pm b$ and
\begin{equation}
  {\begin{array}{c}
  x_n \\
  x_m 
  \end{array}}={\begin{array}{c}
  (n+\frac{1}{2})\mu_BB-{\Delta} \\
  (m+\frac{1}{2})\mu_BB-{\Delta} 
  \end{array}},~~{\begin{array}{c}
  y_n \\
  y_m 
  \end{array}}={\begin{array}{c}
  \sqrt{\mu_+^2-x_n^2} \\
  \sqrt{\mu_-^2-x_m^2}
  \end{array}}.\nonumber
  \end{equation}

Typical plots for $m_z~(m_z=M_z/V)$ variation with $1/B$ is shown in Fig.\ref{vary-mz}. It shows the quantum oscillations in the data for $m_z$ whose amplitude increases with the strength of $B$. One can analytically understand such variations as sum of sinusoidal oscillations which we discuss in the next section. 

\section{Effective mass variation and Zeeman Contribution}

Quantum oscillation measurements of Shubnikov de Haas or De-Haas Van Alphen experiments in NLSM systems like $ZrSiS$\cite{zrsis}, $CaAgAs$\cite{caagas} or Dirac material $PtBi_2$\cite{ptbi2} show signatures of light effective masses and high quantum mobilities in electronic carriers. They exhibit two oscillations frequencies coming from electron ($\alpha$) and hole ($\beta$) pockets in the Fermi surfaces. Our model Hamiltonian also indicates two extremal area of cross-sections in the low energy FS (though the number of QO frequencies doubles on considering the Zeeman term). The oscillating magnetization data, when subtracted from the smooth paramagnetic background can be analyzed using Fast Fourier transformation (FFT) and thus the oscillation frequencies are estimated in experiments. Theoretically such behavior can be described using a Lifshitz-Koservich (LK) formula\cite{shoen}. It shows ${\Delta} M$, the fluctuation in magnetization (such that ${\Delta} M=M-M_{average}$) as sum of oscillating functions each corresponding to a QO frequency F as
\begin{eqnarray}
  {\Delta} M\propto -B^{1/2}R_TR_DR_Ssin[2\pi(\frac{F}{B}-\gamma-\delta)].
  \label{lk}
\end{eqnarray}
This depends on the effective mass $m^*$ through the thermal damping factor $R_T=(\alpha Tm_r/B)/sinh(\alpha Tm_r/B)],~R_D=exp(-\alpha T_Dm_r/B)$ and $R_S=cos(\pi gm_r/2)$ where $\alpha=2\pi^2k_Bm/(\hbar c),~m_r=m^*/m$ and $T_D$ is called the Dingle temperature\cite{zrsis,caagas,ptbi2}. Besides, the first phase term $\gamma=1/2-\phi_B/2\pi$ with $\phi_B$ being the Berry phase and the second phase term $\delta=\pm1/8$ in 3D\cite{zrsis,li}.
  {Interestingly, the spin reduction factor $R_S$ becomes zero when $gm_r$ is an odd integer which is known as the spin-zero effect\cite{s01,s02}. There the QO ceases to exist. So ideally for anisotropic $g$ and $m^\star$, the values of which depend on the field orientations, one can rotate the field and get hold of situations where $R_S$ reduces to become zero and then change its sign resulting in a $\pi$ phase shift in the oscillations\cite{comment}.}
  Usually $m^*$ is obtained via fitting the thermal damping factor $R_T$ of the LK formula with the experimentally obtained variation of the FFT amplitudes with temperature. At zero temperature and considering infinite relaxation time for electrons, one can write, for multiple QO frequencies, ${\Delta} M=B^{1/2}\sum_iC_i(m_r)sin[2\pi(\frac{F_i}{B}-\gamma_i-\delta_i)]$, where the prefactors $C_i$ depend on $m_r$. In our calculations later, we artificially vary $m^*$ to see its effects on the spectral results as well.
\begin{figure}
\includegraphics[width=.65\linewidth]{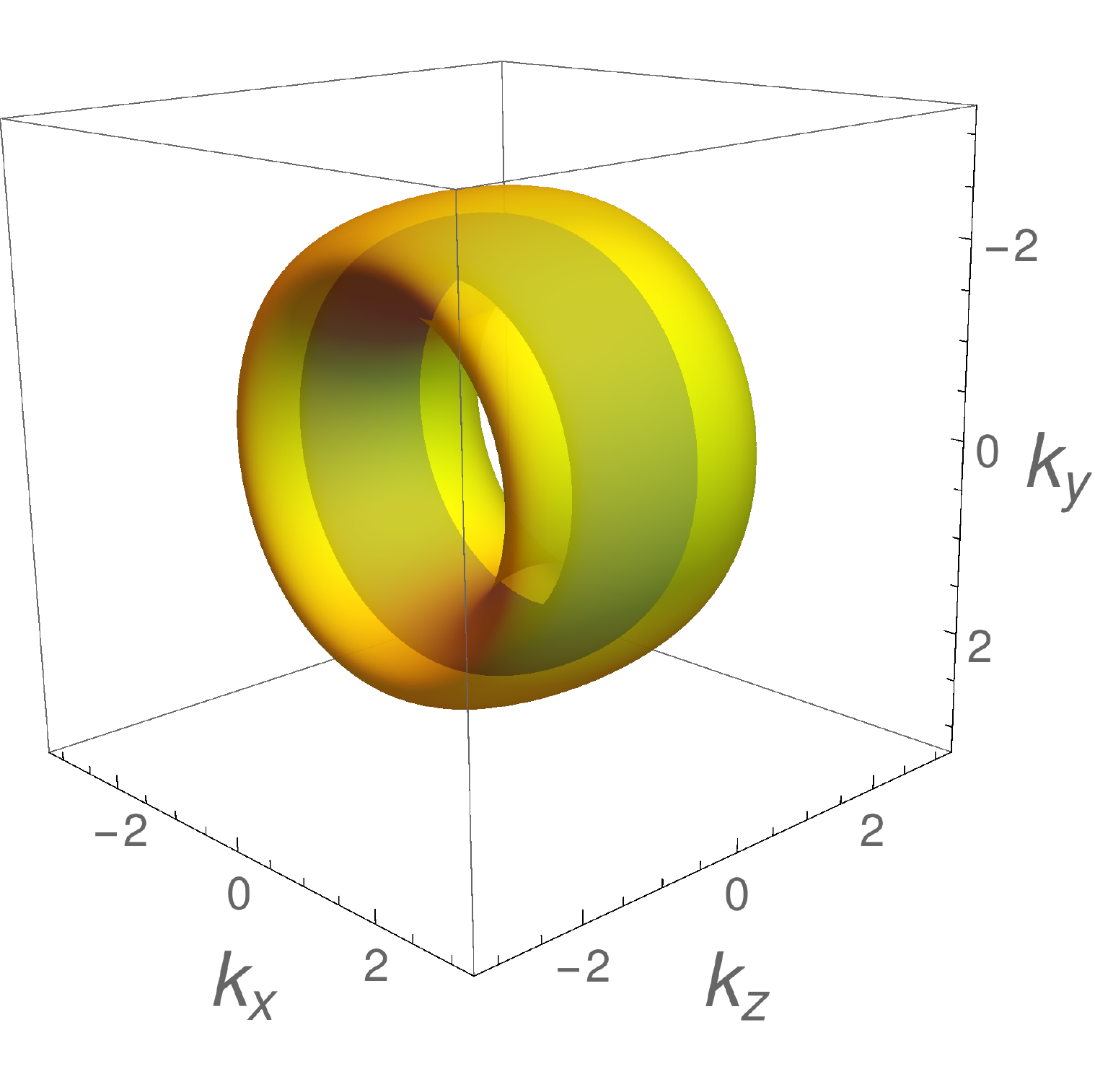}
\caption{Typical Fermi surfaces in presence of the Zeeman term. Generally two distinct surfaces (for $\uparrow$ and $\downarrow$ electrons respectively) give 4 extremal surfaces and hence 4 QO frequencies.} 
\label{fs}
\end{figure}
  
  Now let us get estimates/expressions of these QO frequencies for our NLSM system. Without the Zeeman term, the toroidal Fermi surface gives two extremal surfaces for nonzero $\mu$ (with $\mu<{\Delta}$) values in the $k_z=0$ plane with magnetic field along the $z$ direction. These will be given by
  \begin{equation}
    A^\pm_{ex}=\frac{2\pi m^*}{\hbar^2}({\Delta}\pm\mu)
  \end{equation}
  and following Onsager's relation it leads to two fundamental QO frequencies
  $F^\pm=\frac{\hbar}{2\pi e}A^\pm_{ex}=\frac{m^*}{\hbar e}({\Delta}\pm\mu)$.
  But as we turn on the Zeeman term, one gets two toroidal Fermi surfaces (see Fig.\ref{fs}) at low energies and we need to consider two pairs of frequencies which can be designated as
  \begin{equation}
    F^\pm_\uparrow=\frac{m^*}{\hbar e}({\Delta}\pm(\mu+b))~~{\&}~~F^\pm_\downarrow=\frac{m^*}{\hbar e}({\Delta}\pm(\mu-b))
    \label{frq}
  \end{equation}
  respectively. This is however valid as long as ${\Delta}>\mu+b$ and $b$ small. For $b<\mu$ and ${\Delta}<\mu+b$, there will be 3 QO frequencies. But for $b>\mu$, we need to consider 2 (for ${\Delta}>\mu+b$) or 1 (for ${\Delta}<\mu+b$) QO frequencies.
  {Interestingly, as one plugs in these frequencies of Eq.\ref{frq} in the LK  formula, a few algebric steps finally leads to oscillations with frequencies $F^\pm$ alone but now with an additional prefactor of $~2cos(\pi g\mu_B)$. Thus the FFT spectrum still shows two peaks corresponding to $F^\pm$. We should here mention that these conclusions rely on Zeeman splitting that is linear in $B$. Scenerio changes if nonlinear Zeeman splitting is considered that usually comes for relativistic bands\cite{wang}. However, such detailed analysis is not pursued in the present work. } 
\begin{figure}[t]
\includegraphics[width=\linewidth,height=2.8 in]{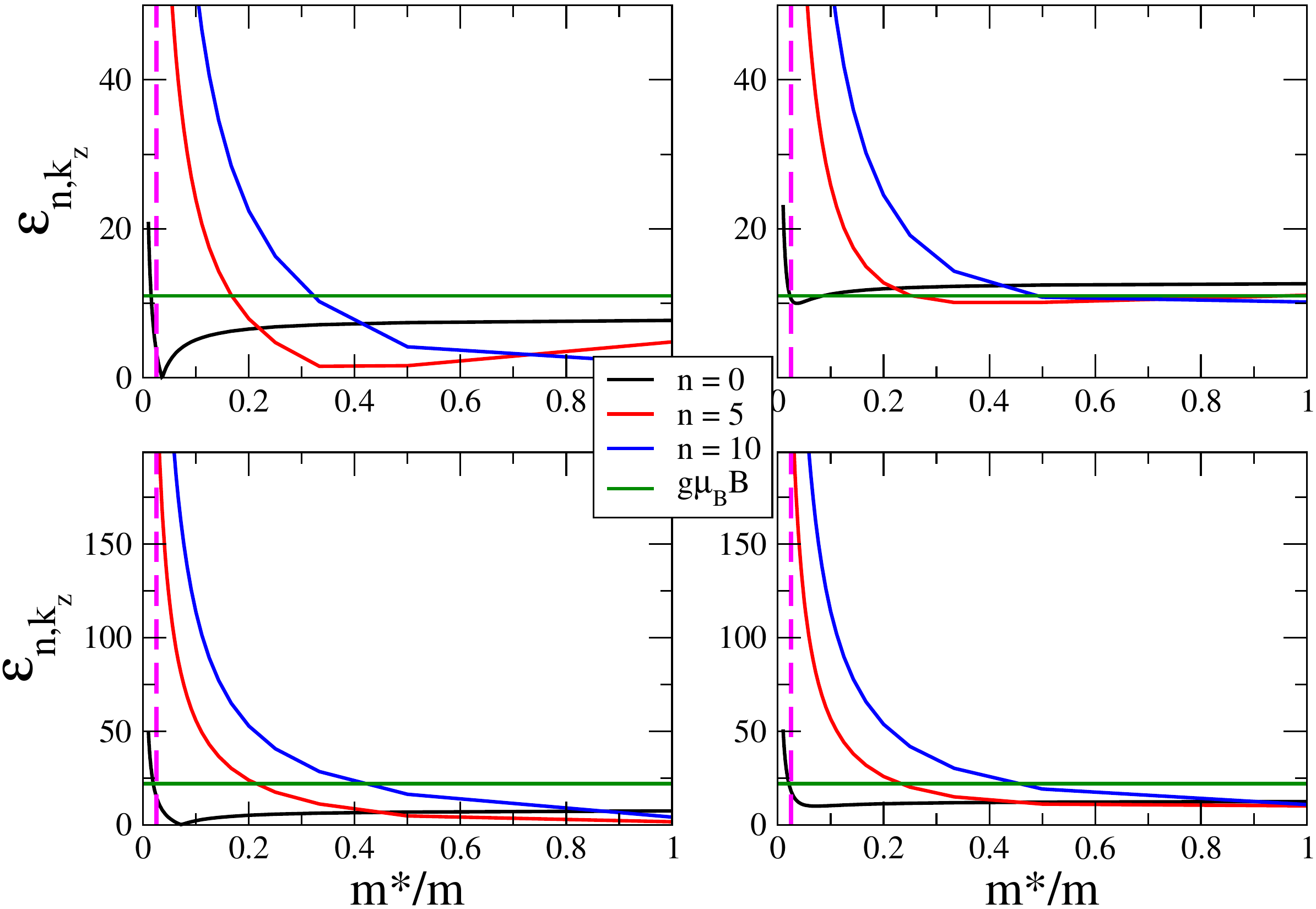}
\caption{(Color online) Energy spectrum for $vp_z=0$ (left) and $vp_z=10$ meV (right) as well as for $B=5T$ (top) and $B=10T$ (bottom) respectively. The Zeeman term with $g=38$ is included and the value of $m_r=0.025$ is highlighted (magenta dashed line) as well.} 
\label{vary-m*}
\end{figure}

 {In this regard we should also clarify that absence of additional FFT peaks due to Zeeman term does not imply absence of Zeeman splitting which will always be there whenever variations with energy is considered. In fact, the joint effect of LL quantization and Zeeman splitting is one important feature of this problem\cite{jeon} and we demonstrate that later in section V and VI from the density of state plots (Fig.6 and Fig.8) for different $m^\star$ values and field directions. Moreover,  at high fields and at low temperatures strong Zeeman splitting is realized during quantum oscillations in these systems.
 Such high field splitting can be realized from Eq.\ref{eq-en}.
  According to LK formula, QO amplitude increases with field magnitude and thus the Zeeman splitting of QO peaks is easily discernible for large magnetic fields\cite{zrsis}.}
  Now we should remember that the Lande $g$ factor for an electron in a band gets modified from its free electron value of $g=2$ due to spin-orbit couplings.
In the present problem, this gets enlarged to very high values depending on the effective mass $m^*$ (see Ref.\onlinecite{g-large}). This in turn amplifies the Zeeman splitting of the LL spectra. Experimentally $g$ can be calculated as $\frac{g}{2}\frac{m^*}{m}=F(\frac{1}{B^+}-\frac{1}{B^-})$ where $B^\pm$ denote two fields at the position of the split peaks in the magnetization plot\cite{zrsis}.

\begin{figure}
\includegraphics[width=\linewidth,height=2.8 in]{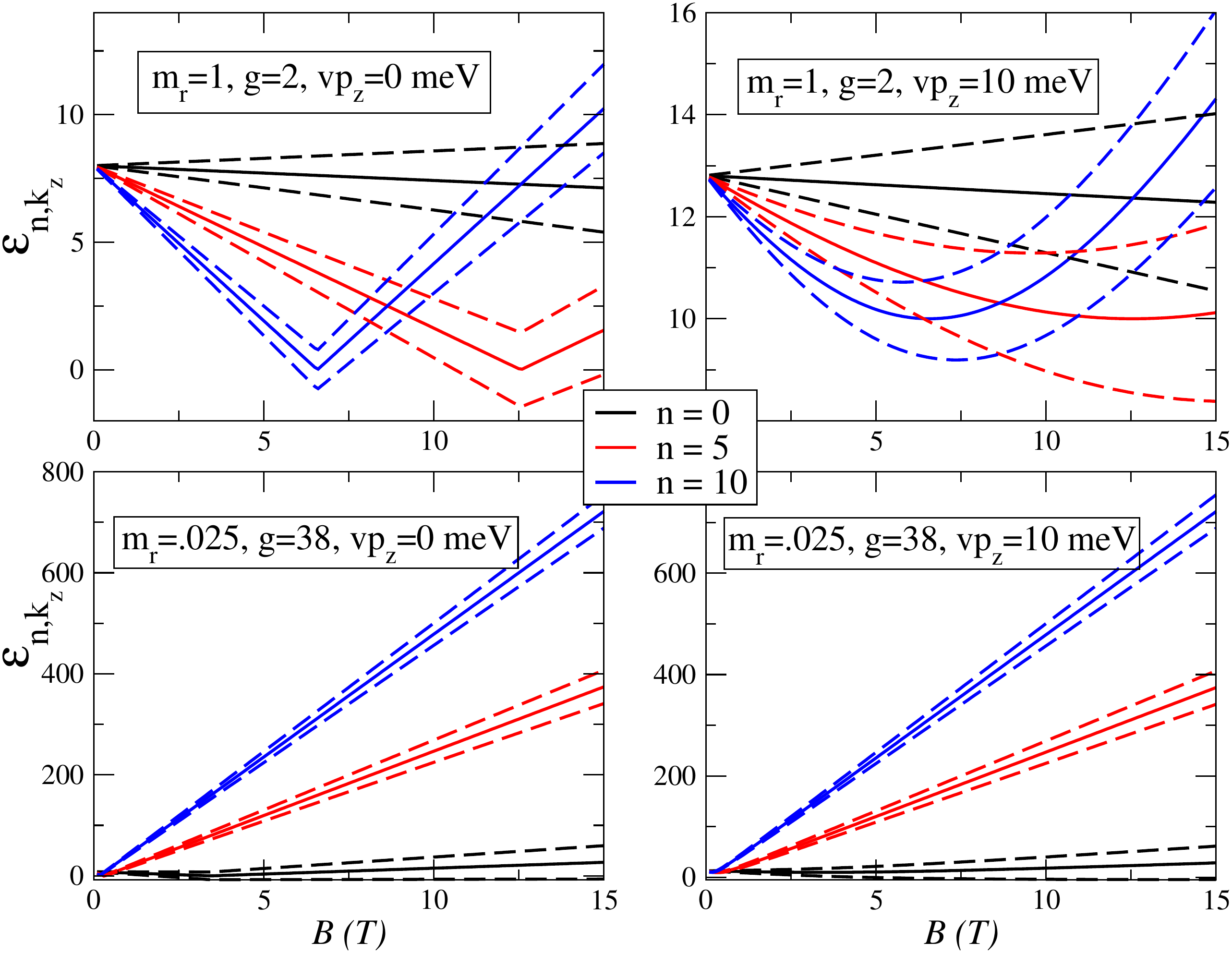}
\caption{Variation of energy spectrum for $vp_z=0$ (left) and $vp_z=10$ meV (right) with magnetic field $B$ both with (dashed lines) and without (solid lines) the Zeeman term.} 
\label{vary-B}
\end{figure}

Here we will use $g=38$ and $m^*=0.025m$, following the findings for NLSM compound $ZrSiS$\cite{zrsis} in order to analyze the effect of Zeeman effect on the NLSM spectra. 
From Eq.\ref{eq-en}, we find that the spectra varies with $m^*$ and it will be interesting to compare such variation with the Zeeman splitting (which is $2b$) to determine the importance of the latter.
In Fig.\ref{vary-m*}, we show such variations which indicate huge $m^*$ dependence on LL dispersions. We find that the Zeeman contribution remains significant in the free electron limit with $m^*=m$, but becomes insignificant in the massless limit $m^*\rightarrow0$. For small masses like $m_r=0.025$, as observed in $ZrSiS$\cite{zrsis}, Zeeman term remains significant only for a few LL such as $n=0$ (see Fig.\ref{vary-m*}).
The Zeeman term becomes important whenever the energy gap is reduced considerably ($i.e.,~gap\le 2b$). It depends not only on $m^*$ but also on $n$ values. Notice that for the other modes with $n=5,10$, the Zeeman contribution is negligible for very small $m^*$ values. For $p_z\ne0$ the Zeeman term becomes even less important as the lower cut-off of energy magnitude becomes nonzero as well. One can also see such behavior from field dependence of the LL spectra as shown in Fig.\ref{vary-B} where the smallness of Zeeman contribution is clearly seen for the small effective mass of $m^*=0.025m$ (even after an enlarged $g=38$ is considered).

Thus for NLSM systems, $e.g.,$ in the compound $ZrSiS$, the combination of effective mass reduction and lande-g factor increase causes the Zeeman splitting to often become comparable to the LL energy spacings.
However, the situation differs for different kind of systems. For example, in GaAs-based heterostructures\cite{fazekas}, the effective mass is taken to be $m^\star\sim0.07m_e$ which increases the cyclotron frequency almost tenfold. On the other hand, Zeeman term also get reduced with $g\sim0.44$. So Zeeman splitting is very small fraction of the LL spacings and its contribution can be neglected. {However for some moderate $g$ values, as considered for example in Dirac semimetallic $PtBi_{2-x}~(x\sim0.4)$ single crystals\cite{xing}, Zeeman splitting will not be completely ignorable.}


\section{Density of states}

Unlike the 3D free electron gas where the Hamiltonian can be decoupled to two independent motions along and perpendicular to the field direction, our present problem constitutes a two level system where the dispersion, $i.e.$, Eq.\ref{eq-en}, is not just sum of contributions from motions parallel and perpendicular to field directions. Thus energy convolution\cite{bennett} can not be utilized to obtain the density of states (DOS). Rather it is obtained from the basic definition as
\begin{eqnarray}
  \rho(E)=\mathcal{N}\sum_{n,k_z} \delta(E-\epsilon_{n,k_z})
  \label{eq10}
\end{eqnarray}
where $\mathcal{N}=eBA/h$ denotes the degeneracy factor, $A$ being the area of the sample in the $xy$ plane ($i.e.,$ plane normal to the field direction).
Numerically one can obtain $\rho(E)$ using Lorentzian approximation of the delta function as
\begin{eqnarray}
  \delta(E-\epsilon_{n,k_z})=Lt_{\eta\rightarrow 0}\frac{1}{\pi}\frac{\eta}{(E-\epsilon_{n,k_z})^2+\eta^2} 
  \label{eq11}
\end{eqnarray}
\begin{figure}[t]
  \includegraphics[width=\linewidth,height=3.5 in]{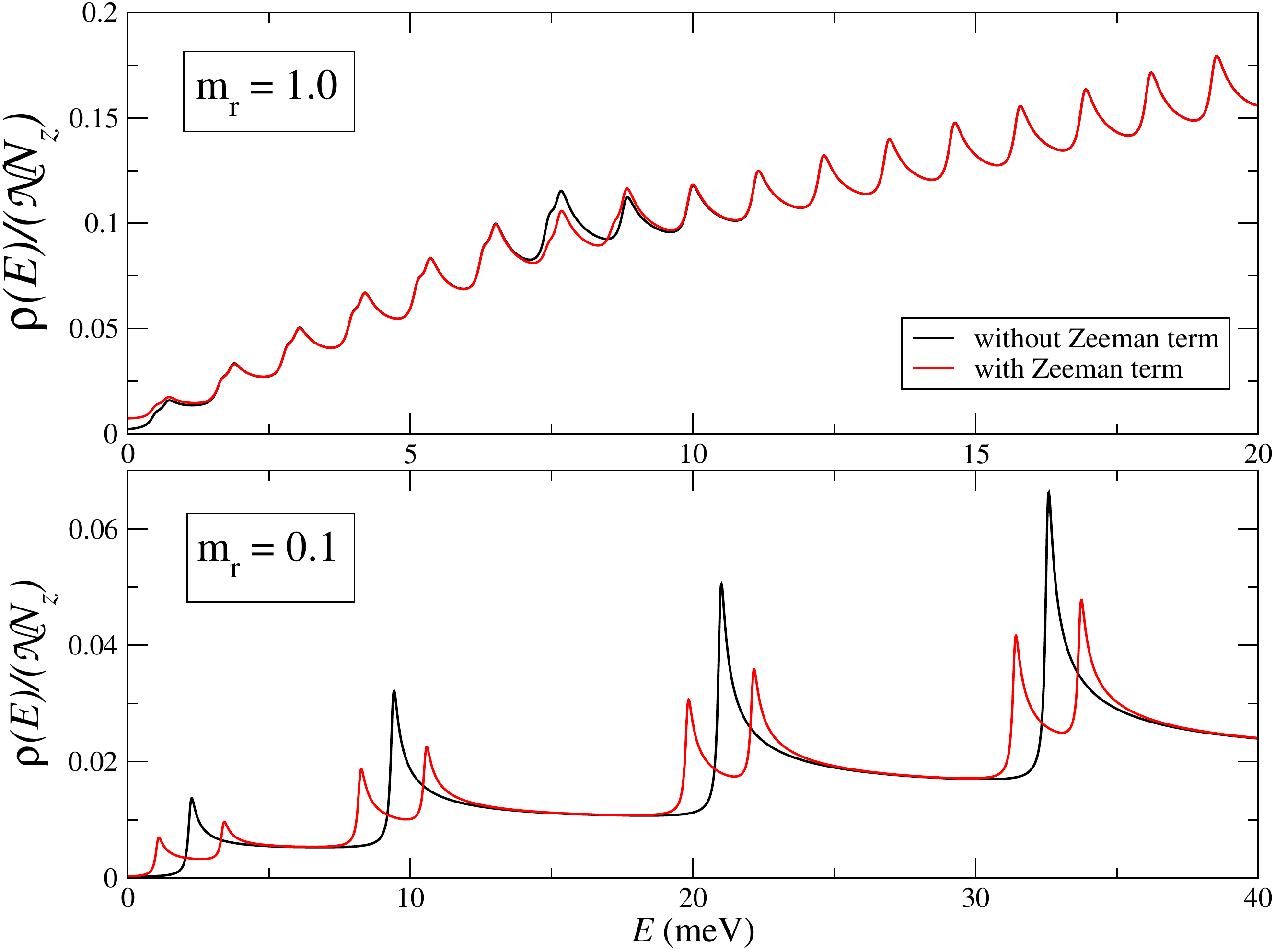}
   \caption{DOS plots for $B=10T$ and ${\Delta}=8 meV$ with both with and without the Zeeman term included for $m_r=1$ (top) and $m_r=0.1$ (bottom). $g=2$ is considered for both the plots.}
\label{dos}
\end{figure}
Furthermore, a sum for Landau levels upto $n=100$ is considered and using periodic boundary conditions we take $k_z=\frac{2\pi}{N_za_0}\nu$ where $N_z$ denotes the total number of $k_z$ points along $z$ direction, $\nu$ is a non-negative integer with $\nu\le N_z$ and $a_0$ is the separation between successive $k$ points along $z$.
In Fig.\ref{dos}, we show the DOS plots for $m^*=m$ and $m^*=m/10$ respectively for $B=10T$ and ${\Delta}= 8$ meV. Notice that the peak positions in DOS almost retains their peak positions as the Zeeman term is turned on for $m_r=1$. This is because the Zeeman splittings become same as the Landau level spacings. On the other hand, for a smaller $m_r=0.1$ one can see the differences in peak positions between DOS plots with and without the Zeeman term as the LL spacings becomes much wider compared to the Zeeman splitting.


\section{Field Parallel to the NODAL plane}

Magnetic oscillations at low energies become topological when the magnetic field directs parallel to the nodal plane and hence this is important in its own right.
The Hamiltonian for the NLSM in presence of field parallel to ${\hat x}$ direction can be given as
\begin{eqnarray}
  H=(\frac{\hbar^2k_\perp^2}{2m^\star}-{\Delta})\sigma^z\tau^0+vp_z\sigma^y\tau^0+b\sigma^0\tau^x .
  \label{eq12}
\end{eqnarray}

We consider the Magnetic field to be ${\bf B}=(B,0,0)$ and the vector potential $A=(0,-Bz,0)$.
The Hamiltonian gets modified via Peierls substitution ${\bf p'=p-eA}$, to become:
\begin{eqnarray}
  H=[\frac{p_x^2}{2m^*}+\frac{(p_y+eBz)^2}{2m^*}-{\Delta}]\sigma^z\tau^0+vp_z\sigma^y\tau^0+b\sigma^0\tau^x
  \label{eq13}
\end{eqnarray}

Now, we introduce the variable,
\begin{eqnarray}
  \tilde{z}=(z+\frac{p_y}{eB})
  \label{eq14}
\end{eqnarray}
that transforms Eq.\ref{eq13} as
\begin{eqnarray}
  H=[\frac{m^*\omega_c^2\tilde{z}^2}{2}-{\Delta}']\sigma^z\tau^0+vp_z\sigma^y\tau^0+b\sigma^0\tau^x
  \label{eq15}
\end{eqnarray}
where ${\Delta}'={\Delta}-\frac{p_x^2}{2m}$ and cyclotron frequency $\omega_c=eB/m^*$.
This Hamiltonian can be numerically diagonalized to get the energy eigenvalues and one can obtain various quantities including density of states from there.
\begin{figure}[t]
  \includegraphics[width=\linewidth,height=3.5 in]{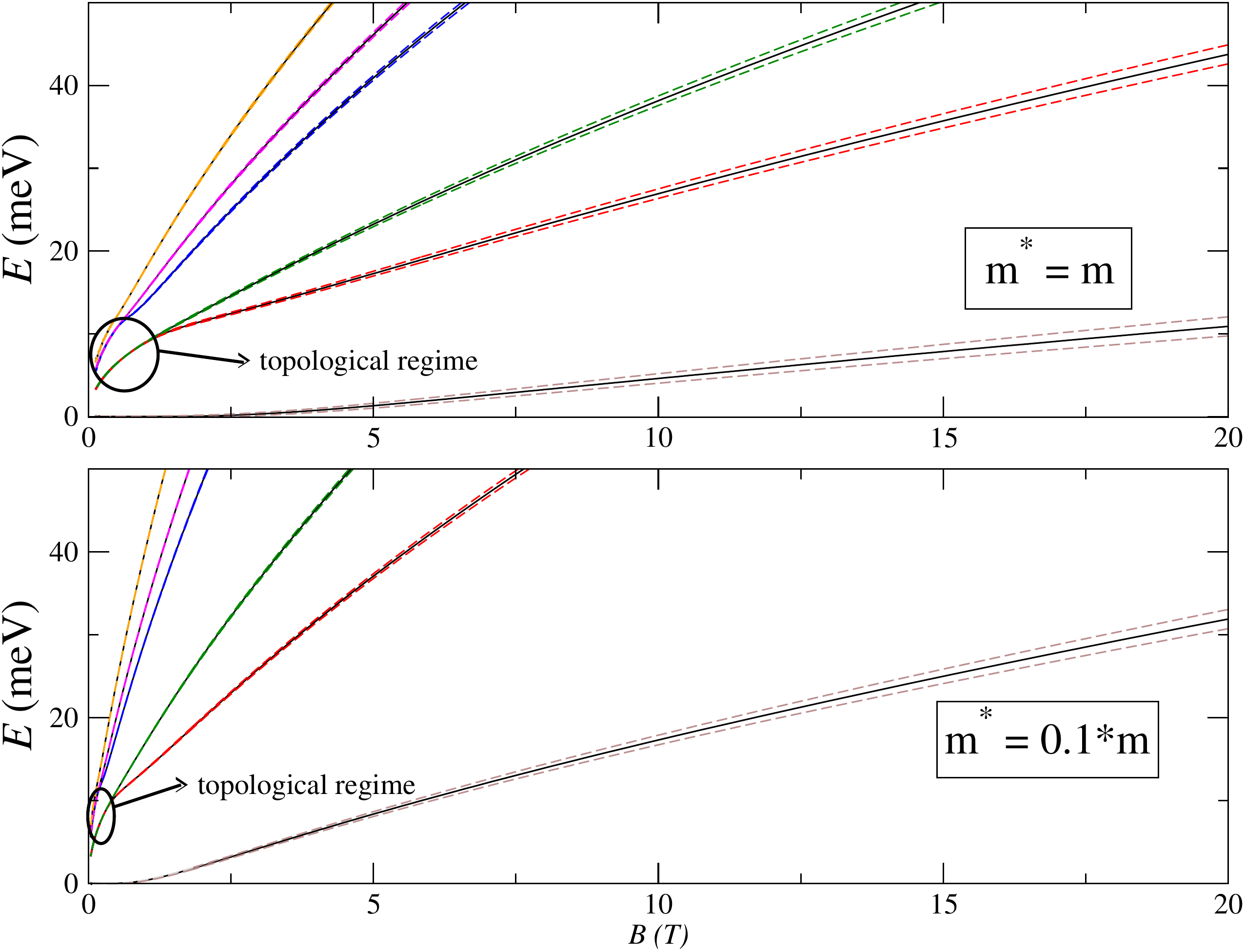}
  \caption{Dispersions in the $p_x=0$ plane in presence (dashed lines) /absence (solid lines) of Zeeman term with effective mass $m^*=m$ (top) and $m^*=0.1*m$ (bottom).}
\label{enBx}
\end{figure}
\begin{figure}[t]
  \includegraphics[width=\linewidth,height=3.5 in]{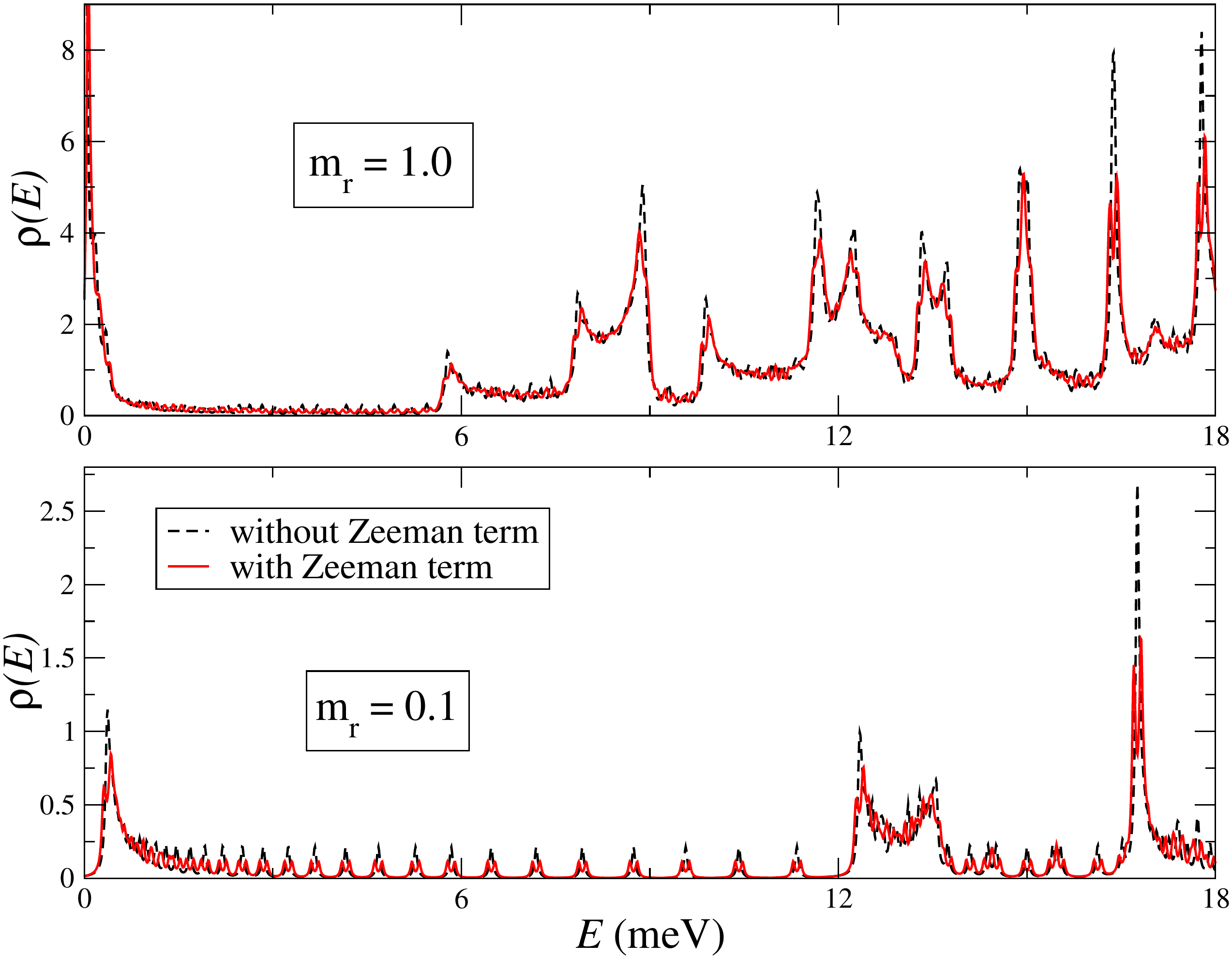}
  \caption{Low energy DOS plots for $B=1T$ and ${\Delta}=12 meV$ with both with and without the Zeeman term included for $m_r=1$ (top) and $m_r=0.1$ (bottom). Here $g=2$ is considered for both the plots.}
\label{dosBx}
\end{figure}

One can get a dimensionless form of the Hamiltonian by defining parameters\cite{kar2}
$\delta=(\frac{2{\Delta}'^3}{m\hbar^2\omega_c^2v^2})^\frac{1}{3}$, $\alpha=(\frac{2v}{m\omega_c^2\sqrt\hbar})^\frac{1}{3}$, $Z=\frac{\tilde{z}}{\alpha\sqrt{\hbar}}$, $P=\frac{\alpha p_z}{\sqrt{\hbar}}$, and $C=\frac{2b}{\alpha^2m\omega_c^2\hbar}$.
This results in the following restructuring of the Hamiltonian:
\begin{eqnarray}
  H=[\frac{m^*\hbar^2\omega_c^2v^2}{2}]^{\frac{1}{3}}[(Z^2-\delta)\sigma^z\tau^0+P\sigma^y\tau^0+C\sigma^0\tau^x]
  \label{eq17}
\end{eqnarray}
which gives the eigenvalues as
\begin{eqnarray}
E=(\frac{m^*\hbar^2\omega_c^2v^2}{2})^\frac{1}{3}[\pm\sqrt{(Z^2-\delta)^2+P^2} \pm C]  .
  \label{eq18}
\end{eqnarray}
Typical positive spectra for $p_x=0$ (where Landau tubes can cross the Fermi surfaces extremally) are shown in Fig.\ref{enBx} in presence/absence of the Zeeman term. Unlike the case with ${\bf B}=B\hat z$, here the Zeeman splitting does not put enough impact as the dispersion energy levels in the present case are comparatively more spread out. {Notice that} the low energy dispersion branches are doubly degenerate which split into two nondegenrate bands at energies $E\sim {\Delta}$ via topological transitions\cite{lim2,cortijo,kar2}. {This is because there is a topological change in the FS at energy $E=\Delta$. For $E>\Delta$, the FS is a spindle torus and the electronic cyclotron loops do not correspond to nonzero Berry phases. But for smaller energies, the FS is a ring torus and it gives a pair of cyclotron loops (at the extremal cross-section of the FS) which individually correspond to nonzero Berry phases resulting in topological oscillations. See Ref.\onlinecite{lim2,cortijo,kar2} for details.} Such low energy topological regime become narrower for smaller effective masses (see Fig.\ref{enBx}). Thus nontrivial oscillations can be observed only for small values of $B$ alone, more so if the NLSM compounds possess small $m^*$ values for the band electrons. Notice that this resizing of the topological regime is decided by the magnitude of the prefactor in Eq.\ref{eq18}. Thus the narrow topological regime can again be streched if we consider smaller values of $v$, the Fermi velocity of electrons along $z$ direction ($i.e.,$ direction normal to nodal plane) , which can be established by considering larger effective mass along $z$ direction.

Next we probe the DOS only at low energies. The DOS indicates occupation at zero energies. In fact that is the most prominent peak in the DOS. At low energies, one can find few other discrete peaks though they don't appear at fixed energy intervals like that is observed corresponding to the multiple of cyclotron frequencies for the field acting perpendicular to nodal plane ($i.e., {\bf B}=B\hat z$).
This is because for ${\bf B}=B\hat x$, the $xy$ planar symmetry is lost as can be seen from the Hamiltonian. One can at most say that for motion perpendicular to the field direction, $i.e.,$ for motion in the $yz$ plane, the low energy spectra can be written as $E_n(p_x)=f(p_x)\sqrt{n}$, $f(p_x)$ being some function of $p_x$ and $n$ a positive integer\cite{B@x}. Moreover, as we consider a smaller $m^*$ instead, a number of characteristic changes occur in the DOS. Most of the weights shift to higher energies. The zero energy peak moves to higher values with much reduced probability of occupation. One gets a series of low intensity low energy peaks, similar to that are seen for $m^*=m$ as well but the major peaks are obtained only for $E>{\Delta}$. These are again the outcome of spreading out of the spectra due to smaller $m^*$ values in compatible with Eq.\ref{eq18}. With Zeeman coupling, all the peaks are bifurcated and these can be seen in the DOS plots of the Fig.\ref{dosBx}.

\section{Summary}
In this paper, we have gone a long way to describe what happens in a simple type I NLSM model as a magnetic field is applied on it and its magnitude is varied through. We quantify the the fluctuation in the Fermi level with the variation in the field strength which increases for larger field values. Then we demonstrate the quantum oscillation in parallel magnetization and its periodicity as well as decay with $1/B$. We quantify how the QO frequencies get modified in presence of Zeeman couplings. The effect of effective mass reduction and Zeeman splitting is vividly described both for field perpendicular to the nodal plane as well as for field parallel to it. Topological oscillations are reported in the latter case for small energies\cite{lim2,cortijo,kar2}. We find such topological regime to become narrower as the effective mass is reduced. The DOS features are examined to see how they are different for two orthogonal directions of the field applied.

Our work can give useful guide on what to expect from the LL spectra, DOS and quantum oscillations in presence of Zeeman splitting, reduced effective masses and enlarged Lande-g factors. Moreover, the Fermi surface structures and QO frequencies can help understanding the FFT spectra obtained from NLSM compounds during quantum oscillations. As a continuation of the present work, one can also study the inter LL transfers of electrons or the defect productions\cite{lz} once periodic variation of the magnetic field\cite{kar2} is considered in presence of Zeeman coupling and reduced $m^*$ values. And it will be equally interesting to probe the entanglement generation\cite{banasri} in NLSM systems following the Floquet Hamiltonian\cite{eckart} corresponding to such periodic driving involving an oscillating magnetic field. Another quite natural extension of this work will be to look at the similar aspects but using a type II NLSM model and probe the genre of Landau level collapses that appear in those cases\cite{he}. Finally, one can also test the outcomes presented here on optical lattices\cite{song} and utilize them for further engineering.

\section*{Acknowledgements}
{SK thanks L. Balicas, R. Schoenemann, W. Zheng and C. S. Yadav for fruitful discussions.}
This work is financially supported by DST-SERB, Government of India under grant no. SRG/2019/002143.




\begin{thebibliography}{99}
\bibitem{comment} The present article does not analyze the details of $g$ and $m^*$ variation with field orientations but wish to study the spin-zero effect in the light of those variations in some future communications.
\end{thebibliography}
\end{document}